\documentstyle[prl,aps,psfig,twocolumn]{revtex}
\begin{document}
\draft
\flushbottom
\twocolumn[
\hsize\textwidth\columnwidth\hsize\csname @twocolumnfalse\endcsname

\title{The Arbitrary Trajectory Quantization Method}
\author{Debabrata Biswas}
\address{
Theoretical Physics Division \\
Bhabha Atomic Research Centre \\
Mumbai 400 085, INDIA}
\date{\today}
\maketitle
\begin{abstract}
The arbitrary trajectory quantization method (ATQM) is a 
time dependent approach to quasiclassical quantization based on
the {\em approximate} dual relationship that exists between
the quantum energy spectra and classical periodic orbits. 
It has recently been shown however, that, for polygonal billiards,
the periodicity criterion must be relaxed to include  
closed almost-periodic (CAP) orbit families in this relationship.
In light of this result, we reinvestigate the ATQM
and show that at finite energies, a smoothened quasiclassical 
kernel corresponds to the modified formula that includes CAP 
families while the $\delta$ function kernel corresponding
to the periodic orbit formula is recovered as $E \rightarrow \infty$.
Several clarifications are also provided.

\vskip 0.05 in
Published in Phys. Rev. E 63, 016213, \copyright The American
Physical Society
 
\vskip 0.1in
\end{abstract}

]
\narrowtext
\tightenlines

\newcommand{\be}{\begin{equation}}
\newcommand{\ee}{\end{equation}}
\newcommand{\bea}{\begin{eqnarray}}
\newcommand{\eea}{\end{eqnarray}}
\newcommand{\Lop}{{\cal L}}
\newcommand{\DB}[1]{\marginpar{\footnotesize DB: #1}}
\newcommand{\q}{{\bf q}}
\newcommand{\kt}{\tilde{k}}
\newcommand{\Lopn}{\tilde{\Lop}}

\section{Introduction}

Semiclassical quantization methods developed over the past three
decades generally rely on the energy domain approach. 
The spectral density, $\rho(E) = \sum_n \delta (E - E_n)$
is thus expressed in 
terms of periodic orbits of the underlying classical dynamics 
and this is commonly referred to as the Gutzwiller trace 
formula \cite{gutzwiller}.
The crossover from the time domain to the energy domain and the
trace formula necessitates
several approximations in the form of stationary phase integrations and
one of these selects periodic orbits as the sole classical ingredient.
The recipe works well in most cases since closed non-periodic
orbits contribute with a lesser weight and can be included only
as a correction. There are instances however when closed non-periodic
orbits contribute with weights comparable to periodic orbits
and the ``periodicity criterion'' must then be relaxed to accommodate 
them \cite{harel_uzy,DB99_2}.
Polygonal billiards provide such an example and perhaps
hold the key to generic intermittent behaviour. 
In these systems, a slight change in the internal angles
results in the destruction of periodic orbit families thereby giving rise
to closed almost-periodic (CAP) families of orbits \cite{explain_CAP}.
Since this is true for {\em each} periodic family in {\em every}
neighbouring polygon, CAP families actually outnumber periodic
families of any given polygon. Further, they contribute with
weights comparable to periodic families in the semiclassical 
trace formula and are hence indispensable for semiclassical
quantization. This makes the energy domain approach rather
cumbersome to implement and a time domain approach is 
thus preferable.

For polygonal billiards, there are two available 
approaches in the time domain. The
first of these developed by Heller and Tomsovic \cite{heller_td} relies on
a semiclassical construction of the time dependent propagator
to evaluate the auto-correlation function, $C(t) = <\psi|\psi(t)>$ 
where $\psi(\q,t) = <\q|\psi(t)> = 
\sum_n c_n \phi_n(\q) e^{-iE_nt/\hbar} \simeq \int 
K_{s.c}(\q,\q',t) \psi(\q',0) d\q'$. Here
$\{\phi_n\}$, $\{E_n\}$ and $\{c_n\}$ are the eigenfunctions,
eigenvalues and expansion coefficients respectively while 
$K_{s.c}(\q,\q',t)$ is the semiclassical propagator 
constructed using classical trajectories joining 
a pair of points $(\q,\q')$ at each time $t$.
A power spectrum of $\psi(\q,t)$ thus yields the quantum eigenvalues.
The method has been
successfully applied to the stadium billiard  but to the best of our
knowledge, it has not been used for quantizing polygonal billiards.

\begin{figure}[tbp]
{\hspace*{0.05cm}\psfig{figure=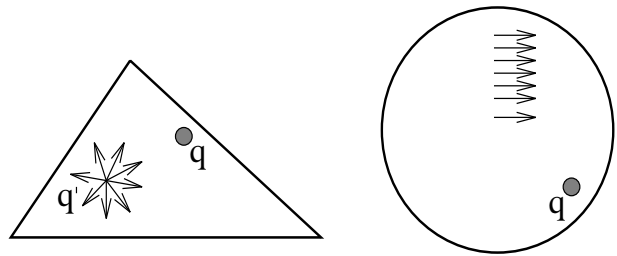,height=3.25cm,width=8cm,angle=0}}
{\vspace*{.13in}}
\caption[ty]{The arbitrary trajectory quantization recipe. In case
of the polygon, trajectories with different linear momentum span the
entire constant energy surface. In case of the circle, this is
achieved by shooting trajectories with different angular momentum.
In both cases, the (weighted) fraction of trajectories in the cell
around $\q$ is recorded.}
\label{fig:1}
\end{figure}

The second approach \cite{PRL97,93_dec} is simpler to adopt 
and relies on classical
propagation (or its quasiclassical \cite{semi_quasi} 
adaptation where necessary).
In case of a polygon, it involves shooting arbitrary trajectories in 
various directions from a point (call it $\q'$ - see fig.~\ref{fig:1}, 
left) and at each time
step, recording the (weighted) fraction, $F(t)$ of trajectories 
that are in an $\epsilon$ neighbourhood of a point $\q$. 
The peaks in the power spectrum of $F(t)$ are then related 
to the quantum eigenvalues.    

Apart from the simplicity, 
the arbitrary trajectory quantization method 
(ATQM) (see \cite{PRL97}) is perhaps the only semiclassical 
scheme that has been successful in determining the first few 
quantal energy levels of generic polygonal billiards
\cite{shudo}.
However, the theory as presented in 
\cite{PRL97} leaves a number of questions unanswered. For one,
the role of the $\epsilon$ neighbourhood is unclear. Besides, as
the theory uses periodic orbit quantization to relate the peaks 
in the power spectrum of $F(t)$ with the quantum eigenenergies,
the role of CAP trajectories must be clarified.
We shall thus reinvestigate
ATQM from this viewpoint and understand why it works.

The plan of the paper is as follows. In section \ref{sec:review},
we shall recapitulate the existing work on the arbitrary
trajectory quantization method. 
The modifications that we shall carry out to
account for the inclusion of closed almost-periodic orbits can be
found in section \ref{sec:new} and this constitutes the main
part of this paper. Finally, a discussion on errors and a 
summary of our results can be found in section \ref{sec:conc}. 
 
\section{Arbitrary trajectory quantization : a review}
\label{sec:review}

The arbitrary trajectory quantization method relies on a suitable
(quasiclassical) adaptation of the classical evolution operator
that propagates a density under a flow that we denote by $q^t$. 
For polygonal billiards, the flow occurs on an 
invariant surface  that is 2-dimensional and
characterized by the two constants of motion, $\{E,\varphi\}$
where $E$ is the energy and $\varphi$ denotes the second constant.
The surface has the topology of a sphere with $g$ (called the genus)
holes where $g$ can be determined from the internal angles of the 
polygon. The motion on the invariant surface can alternately be
viewed on a singly connected surface obtained by executing $2g$
cuts and with edges appropriately identified.

\par
\vbox{
\begin{figure}[tbp]
{\hspace*{-1.5cm}\psfig{figure=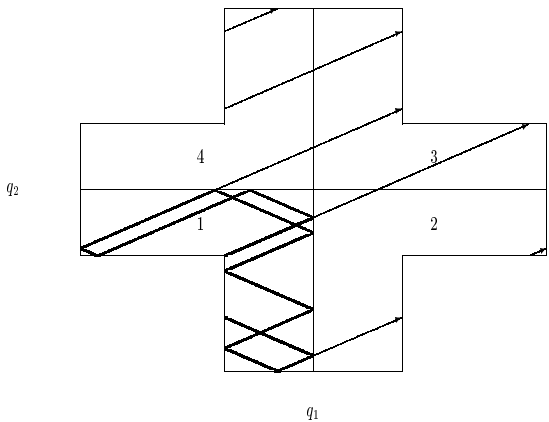,height=6.5cm,width=13cm,angle=0}}
\caption[ty]{The singly connected region for an L-shaped billiard consists
of four copies with edges appropriately identified. A trajectory originating
near the $3\pi/2$ vertex in $1$ is plotted in configuration space using bold
lines and the corresponding unfolded trajectory is also shown.
The latter consists of parallel segments and the trajectory can be 
parameterized by the angle $\varphi$ that it makes for example with 
the $q_1$ axis.}
\label{fig:2}
\end{figure}
}

As a trivial example,
consider the rectangular billiard. 
The singly connected region is a larger rectangle consisting
of four copies corresponding
to the four directions that a trajectory can have and these
can be glued appropriately to form a torus. 
As a non-trivial
example, consider the L-shaped billiard of fig.~\ref{fig:2} which is
pseudo-integrable with its invariant surface having, $g=2$.
Alternately, the surface can be represented by a singly connected
region in the plane and consists of four copies
corresponding to the four possible directions an orbit can
have and these are glued appropriately.
A trajectory in phase space thus consists of parallel segments at an 
angle $\varphi$ \cite{circle}
measured for example with respect to one of the sides.
It will be useful to note at this point that the same trajectory 
can also be represented by parallel segments at angles
$\pi - \varphi$, $\pi + \varphi$ and $2\pi - \varphi$.
In general, the number of
directions for representing a trajectory equals the number of copies, $N$,   
that constitute the invariant surface.

\par Before considering the question of quasiclassical
quantization, we first introduce the appropriate
{\it classical} evolution operator. For
integrable systems, this is easily defined as

\be
\Lop^t{\circ}\phi(\theta_1,\theta_2) =
\int d\theta_{1}'d\theta_{2}'\;
\delta (\theta_1 - \theta_{1}'^t)\delta (\theta_2 - \theta_{2}'^t)
\;\phi(\theta_1',\theta_2') \label{eq:prop}
\ee

\noindent
where $\theta_1$ and $\theta_2$ are the angular coordinates on the
torus and evolve in time as $\theta_i^t = \omega_i (I_1,I_2)t +
\theta_i$ with $\omega_i = \partial H(I_1,I_2)/\partial I_i $ and
$I_i = {1\over 2\pi}\oint_ {\Gamma_i} {\bf p.dq} $.
Here $\Gamma _i, i = 1,2$ refer to the
two irreducible circuits on the torus and ${\bf p}$ is the momentum
conjugate to the coordinate ${\bf q}$.

\par On a general two dimensional invariant 
surface parameterized
by $\varphi$, the classical propagator is expressed as \cite{PRL97}

\be
\Lop^t (\varphi) {\circ}\phi({\bf q}) = \int d{\bf q'}\;
\delta({\bf q} - {\bf q}'^t(\varphi))
\;\phi({\bf q}') \label{eq:def1}
\ee

\noindent
where ${\bf q}$ refers to the position in the singly connected region 
(see fig.~\ref{fig:2}) and ${\bf q}'^t(\varphi)$ is the  
time evolution parameterized by $\varphi$ as described above.
We denote by $\{\Lambda_n(t;\varphi)\}$, the eigenvalues
of $\Lop^t(\varphi)$ and from its multiplicative nature, it
follows that $\Lambda_n(t;\varphi) = e^{\lambda_n(\varphi)t}$.
We are interested here in the form of $\lambda_n(\varphi)$
and to this end we shall evaluate

\bea
\int d\varphi~{\rm Tr}~\Lop^t(\varphi) & = &\int d\varphi \sum 
e^{\lambda_n(\varphi)t} \nonumber \\ 
& = & \int d\varphi \int d{\bf q} \;\delta({\bf q} -
{\bf q}^t(\varphi)) \label{eq:trace1}.
\eea

\noindent
The delta function kernel ensures that the only orbits that
contribute are ones that are periodic or close on the invariant surface
after a time $t$. The {\bf q} integrations are thus simpler to perform if we
transform to a local coordinate system with one component
parallel to the trajectory and the other perpendicular.
Thus $\delta_{\|}(q_\| - q_\| ^t) = 
\sum_p \sum_r {1\over v}\delta (t-rT_p) $
where $v$ is the velocity, $T_p$ is the period of the
orbit and $r$ is the repetition number. Similarly, for
an orbit of period $rT_p$ parameterized by the angle
$\varphi_p$, $\delta_{\bot} (q_\bot - q_\bot ^{rT_p}) =
\delta (\varphi - \varphi_p)/{\left |\partial q_\bot/\partial
\varphi \right |_{\varphi = \varphi_p} }$ where 
$\left |\partial q_\bot/\partial \varphi 
\right |_{\varphi = \varphi_p} = rl_p$ for marginally
unstable billiards. Putting these results together and
noting that each periodic orbit occurs in general at
$N_p$ different values of $\varphi$,  we finally have
 
\bea
\int d\varphi \sum 
e^{\lambda_n(\varphi)t} & = & \sum_p \sum_{r=1}^{\infty}
 {a_p N_p\over rl_p}\delta(l-rl_p)  \nonumber \\ 
& \simeq & N \sum_p \sum_{r=1}^{\infty}
 {a_p\over rl_p}\delta(l-rl_p)
\label{eq:trace2}
\eea

\noindent
where $l = tv$ and the summation over $p$ refers to all primitive 
periodic orbit families with length $l_p$ and  occupying an area $a_p$.
Note that in Eq.~(\ref{eq:trace2}), we have replaced $N_p$ by $N$
since for most long orbits, $N_p \simeq N$.
Also, we have neglected the
influence of isolated orbits to simply matters. We shall continue
to make this approximation through the rest of this paper and 
justify its use at the end.

In some cases, it is possible to interpret the periodic
orbit sum in Eq.~(\ref{eq:trace2}) starting with the
semiclassical ($E\rightarrow \infty $) trace formula for marginally
stable systems :

\bea
\rho(E) & \simeq & \rho_{av}(E) \nonumber \\
 & + & {1\over \sqrt{8\pi^3}}
\sum_p \sum_{r=1}^{\infty} {a_p\over \sqrt{krl_p}}\cos(krl_p -
{\pi\over 4} - r\mu_p) \label{eq:semi_trace}.
\eea

\noindent
Here $\rho_{av}(E)$ refers to the average
density of quantal eigenstates, $k = \sqrt{E}$, 
$l_p$ is the length of a primitive 
periodic orbit family, $\mu_p = \pi n_p + \nu_p\pi/2$,
$n_p$ the number of (phase-altering) bounces
that it suffers at the boundary and $\nu_p$ the number of 
caustics encountered by the orbit. Note that in the 
Neumann case, $n_p = 0$ since there is no phase loss 
on reflection while for 
polygonal billiards, $\nu_p = 0$. For convenience, 
we have $\hbar = 1$, $v = 1$ and the mass $m=1/2$.
Starting with the function  

\be
 \sum_n f(\sqrt{E_n}l) e^{-\beta E_n} =
\int_\Delta^\infty dE\; f(\sqrt{E}l) e^{-\beta E} \sum_n \rho(E)
\ee

\noindent
where $f(x) = \sqrt{{2\over \pi x}} \cos(x - \pi/4)$  
and $ 0 < \Delta < E_0 $,
it is possible to show using Eq.~(\ref{eq:semi_trace}) that
for polygonal billiards \cite{prl1} 

\be
\int d\varphi \sum 
e^{\lambda_n(\varphi)t} \simeq 2\pi Nb_0 + 2\pi N \sum_n f(\sqrt{E_n}l) 
\label{eq:trace5}.  
\ee

\noindent
where $\{E_n\}$ are the Neumann eigenvalues of the system and $b_0$ is
a constant \cite{prl1,rap1}.
Thus $\lambda_n(\varphi) = i\sqrt{E_n}\sin(\varphi) \cite{asymp}$.
This is the central result of \cite{PRL97} when the Maslov
phases are zero.

For integrable polygons, Eq.~(\ref{eq:trace5})
can in fact be derived directly starting from Eq.~(\ref{eq:prop}).
The eigenfunctions,
$\{ \phi_n(\theta_1,\theta_2)\}$, on the torus are such that
$\phi_n(\theta_1^t,\theta_2^t) = \Lambda_n(t) \phi_n(\theta_1,\theta_2)$
where $\Lambda_n(t) = e^{i\alpha_n t}$. On demanding that
$\phi_n(\theta_1,\theta_2)$
be a single valued function of $(\theta_1,\theta_2)$, it follows
that $\phi_{\bf n}(\theta_1,\theta_2) = e^{i(n_1\theta_1 + n_2\theta_2)}$
where ${\bf n} = (n_1,n_2)$  is a point on the integer lattice.
Thus the eigenvalue,
$\Lambda_{\bf n}(t) = {\rm exp}\{it(n_1\omega_1 + n_2\omega_2)\}$.

To illustrate the relationship between $ \int  d\varphi \sum 
e^{\lambda_n(\varphi)t} $ and $\{E_n\}$, consider a rectangular 
billiard for which the
Hamiltonian expressed in terms of the actions,
${I_1,I_2}$ is $H(I_1,I_2) = \pi^2(I_1^2/L_1^2 + I_2^2/L_2^2)$
where $L_1,L_2$ are the lengths of the two sides.
With $I_1 = \sqrt{E}L_1\cos(\varphi )/\pi$ and
$I_2 = \sqrt{E} L_2\sin(\varphi )/\pi$, it is easy to
see that at a given energy, $E$, each torus is parameterized by a
particular value of $\varphi$. Thus 

\be
\Lambda_{\bf n}(t;\varphi) =
e^{i2\pi t\sqrt{E}(n_1\cos(\varphi)/L_1 + n_2\sin(\varphi)/L_2)}
\ee

\noindent
and 

\bea
\int d\varphi \sum 
e^{\lambda_n(\varphi)t} & = & 
\sum_{\bf n} \int_{-\pi - \mu_n}^{\pi - \mu_n} d\varphi\;
e^{il\sqrt{E_{\bf n}}\sin(\varphi + \mu_{\bf n})} \nonumber \\
& = & 2\pi \sum_{\bf n} J_0(\sqrt{E_{\bf n}}l)  \label{eq:bessel}
\eea

\noindent
where $l = 2t\sqrt{E}$, $\tan(\mu_{\bf n}) = n_1L_2/(n_2L_1)$
and $E_n = \pi^2(n_1^2/L_1^2 + n_2^2/L_2^2)$.
On separating out ${\bf n} = (0,0)$ from the rest, restricting
the summation to the first quadrant of the integer lattice and
noting that for a rectangle $b_0 = 1/4$, Eq.~(\ref{eq:trace5})
follows.

\par The classical evolution operator thus serves to determine
the Neumann spectrum in polygonal billiards.
Appropriate modifications however need to be made for the
Dirichlet spectrum or for systems that have caustics
(the circle billiard is an example) and the construction
of the evolution operator is then guided by the nature
of the semiclassical trace formula (Eq.~\ref{eq:semi_trace}).

\par The {\it quasiclassical} evolution operator, 
$\Lop_{qc}$, linking the classical eigenvalues to the 
the desired semiclassical eigenvalues can be defined as 

\be
\Lop_{qc}^t(\varphi){\circ}\phi({\bf q}) = \int d{\bf q'}\;\delta({\bf q} -
{\bf q}'^t(\varphi)) e^{-in(t)\pi - i\nu(t){\pi\over 2}}
\phi({\bf q}') \label{eq:prop1}
\ee

\noindent
where $\nu(t) = \nu({\bf q'}^{t}(\varphi))$
and $n(t) = n({\bf q'}^{t}(\varphi))$ count respectively
the number of caustics and (phase altering)
reflections encountered by the trajectory
${\bf q'}^{t}(\varphi)$ in time $t$.
As before, the multiplicative
nature of $\Lop_{qc}^t(\varphi)$ implies that its spectrum
is of the form $\{e^{\lambda_n(\varphi)t}\}$ and it remains to be
shown that for the quasiclassical operator defined in Eq.~(\ref{eq:prop1}),
$\{\lambda_n\}$ has a one-to-one correspondence with the
appropriate quantum spectrum.

As before, we shall evaluate 

\bea
\int d\varphi~{\rm Tr}~\Lop_{qc}^t & = & 
\int_n d\varphi \sum e^{\lambda_n(\varphi)t} 
\nonumber \\ 
& = &
\int d\varphi \int d{\bf q}\; \delta({\bf q} -
{\bf q}^t(\varphi))\; e^{-in(t)\pi - i\nu(t){\pi\over 2}}
\label{eq:trace6}
\eea

\noindent
The trajectories that contribute are once more periodic due to the delta
function in the kernel. The ${\bf q}$ and $\varphi$ integrations 
can be performed along similar lines and we finally have

\be
\int d\varphi {\rm Tr}~\Lop_{qc}^t  \simeq
 N~\sum_p \sum_{r=1}^{\infty} {a_p  \over rl_p}
\delta(l-rl_p) e^{-irn_p\pi - ir\nu_p{\pi\over 2}} \label{eq:trace7}
\ee

\noindent
Starting with the function $\sum_n\; g(\sqrt{E_n}l)\exp(-\beta E_n)$, it
follows from Eq.~(\ref{eq:semi_trace}) that for $\beta \rightarrow 0^+$,

\be
\int d\varphi~{\rm Tr}~\Lop_{qc}^t = \sum_n \Lambda_n(t) =
2\pi N C + 2\pi N \sum_n g(\sqrt{E_n}l) \label{eq:pre_final}
\ee

\noindent
where $\{E_n\}$ now refers to the desired quantum spectrum,
$g(x) = \sqrt{2 /(\pi x)}~{\rm exp}(-ix + \pi/4)$ and $C$ is a constant.
Since

\be
g(x) \simeq {1\over \pi}\int_0^{2\pi}~e^{-ix\sin(\varphi)}d\varphi
\ee

\noindent
for large $x$, it follows that for $v = 1$,

\be
\lambda_n(\varphi)= i\sqrt{E_n}\sin(\varphi) \label{eq:final}
\ee

\noindent
Eq.~(\ref{eq:final}) forms the central result of \cite{PRL97}.

\begin{figure}[b]
{\hspace*{-1.5cm}\psfig{figure=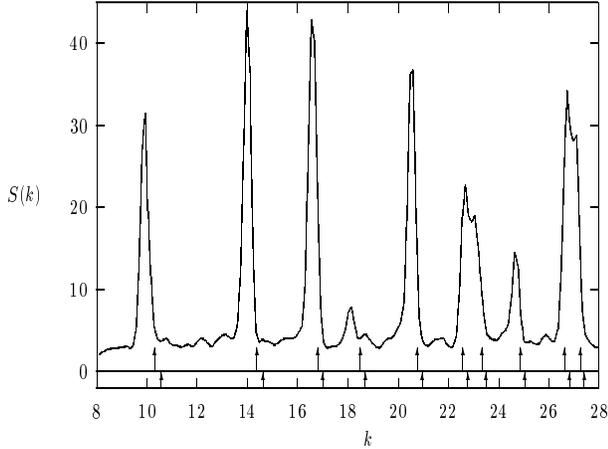,height=7cm,width=14cm,angle=0}}
{\vspace*{.001in}}
\caption[ty]{The power spectrum, $S(k)$
for the triangular enclosure with base length unity.
The top set
of arrows mark the first ten Dirichlet
eigenvalues of the triangle. The bottom set are obtained
from Bogomolny's transfer operator method \cite{bogo92}. 
Note that the non-zero cell size ($\epsilon$) corresponds 
to the hat-function kernel rather than the delta function kernel
used so far. For more details, see text.}
\label{fig:3}
\end{figure}

To demonstrate this result, we shall consider a triangle with
base angles ($\pi/5,3\pi/10$) and evaluate the power spectrum, 
$S(k)$, of the (phase weighted) fraction
of trajectories averaged over 150 cells of radius $\epsilon = 0.03$.
The $\varphi$
integration is performed using 300 trajectories each of which has a
length $2048 \times 0.025$. The result is plotted in fig.~\ref{fig:3}.
The peak positions of $S(k)$ approximate
the exact quantum eigenvalues well. Note however that the 
cells have a non-zero value of $\epsilon$ and the power spectrum 
depends sensitively on the choice of this quantity. In the following 
section therefore, we shall develop an appropriate theory that takes
into account the non-zero $\epsilon$ requirement.

\section{Arbitrary Trajectory Quantization : Modifications}
\label{sec:new} 

As pointed out, 
the existing theory does not require the $\epsilon$
neighbourhood depicted 
in fig.~\ref{fig:1}. All one needs is the expansion of the
kernel 

\be
K(\q,\q'^t(\varphi)) = \sum_n \phi_n(\q) \phi^*_n(\q') 
e^{i\sqrt{E_n}\sin(\varphi)t}
\ee

\noindent
where $\phi_n$ are the 
eigenfunctions of $\Lop_{qc}^t$. The power spectrum, $S(k)$, of
$ \int d\varphi~K(\q,\q'^t(\varphi))$ then
has peaks at $\sqrt{E_n}$. In practice, this implies that
$\epsilon$ must tend to zero since the kernel has a delta
function. The results however get worse
as $\epsilon$ is reduced to zero. A non-zero cell size on the other hand 
gives satisfactory results but is equivalent to a smoothened
kernel \cite{epsilon_arbit}.  
There is thus a gap in the present understanding and it is necessary 
to take a fresh look at the theoretical framework.

The trouble in effect lies with the semiclassical trace 
formula (Eq.~(\ref{eq:semi_trace})) which
neglects the contributions of closed almost-periodic trajectories.
In case of a polygon, these occur in families with the same
symbolic dynamics and across which the 
action varies slowly. To see this, consider an arbitrary triangle 
$T$. In its immediate neighbourhood (obtained by changing the
angles slightly), there exists an infinity 
of triangles $\{T^{(i)}\}$, {\em each with a distinct
periodic orbit spectrum} but having the same symbol sequence (in which 
the sides are visited) for times that depend on the differences in
angles. Assume now that there exists a periodic orbit corresponding 
to a sequence $S_k$ 
for the triangle $T^{(j)}$. Then, for all other
triangles in its neighbourhood, this sequence contributes to the
semiclassical trace formula,
an amount (nearly) equal to the periodic orbit contribution of  $T^{(j)}$
provided $\pi w_i \varphi_e^{(i)}  << \lambda$ \cite{DB99_2}. 
Here $\Delta\theta_i$ is the 
angle between the initial and final momentum of the orbit,
$w_i$ is the transverse extent of the family and
$\lambda$ is the de Broglie wavelength. Thus, corresponding to {\em every}
periodic family in {\em each} of the triangles $\{T^{(i)}\}$, there
exists an almost-periodic family in the triangle $T$ whose
contribution is comparable to that of periodic orbit families in
these neighbouring triangles.

The semiclassical trace formula should thus  
incorporate CAP orbits and the modified expression is \cite{DB99_2} : 

\bea
\rho(E) & \simeq & \rho_{av}(E) + \sum_i  {a_i \over
\sqrt{8\pi^3 k l_i}} \nonumber \\
& \times & \cos(kl_i  -  \pi/4)
{\sin(k\Delta\theta_i w_i/2)\over k\Delta\theta_i w_i/2}
\label{eq:semi_poly_modif}
\eea

\noindent
where the sum over $i$ runs over closed almost-periodic {\em and} periodic
orbit families of transverse extent $w_i$, $l_i$ is the average length
of such a family (taken as the length of the orbit at the centre
of the band) while $\Delta\theta_i$ is the 
angle between the initial and final momentum of the orbit.
Note that at any finite $k$, there exists a  CAP family for which 
the de Broglie wavelength, $\lambda >> \pi w_i \Delta\theta_i$
so that the family contributes to the modified trace formula 
with a weight comparable to that 
of periodic families (${\cal O}(1/k^{1/2})$).

The modified trace formula (Eq.~\ref{eq:semi_poly_modif}) however
fails to relate the quantum eigenvalues with the
eigenvalues of $\Lop_{qc}$ since its $\delta$ function kernel 
chooses only periodic orbits. The kernel function must therefore
be suitably smoothened to achieve such a correspondence.

In order that there exists a direct relationship between the 
eigenvalues of a quasiclassical operator, $\Lopn_{qc}$ and the quantum
eigenvalues, $\{E_n\}$, the kernel function corresponding
to $\Lopn_{qc}$ must be \cite{adhoc}

\be
\tilde{K}_{qc}(\q,\q',t) = K_s(\q,\q',t) e^{-in(t)\pi - i\nu(t){\pi\over 2}} 
\label{eq:prop3}
\ee

\noindent
where 

\be
K_s(\q,\q',t) = 
{\sin\left(\kt(q_\| - {q'_\|}^t(\varphi)\right) \over \pi(q_\| - 
{q'_\|}^t(\varphi))}{\sin\left(\kt(q_\bot - {q'_\bot}^t(\varphi)
\right)\over \pi(q_\bot - {q'_\bot}^t(\varphi))}. \label{eq:prop4}
\ee
 
\noindent
Note that $\lim_{\kt\rightarrow \infty} K_s(\q,\q',t) = \delta(
\q - \q'^t(\varphi))$. The subscript $s$ thus
denotes the smoothened kernel. 
As before, the multiplicative
nature of $\Lopn_{qc}^t(\varphi)$ implies that its spectrum 
is of the form $\{e^{\lambda_n^s(\varphi)t}\}$ and we shall now
show that for the quasiclassical operator defined in 
Eq.~(\ref{eq:prop3}), $\{\lambda_n^s\}$ has a one-to-one 
correspondence with the appropriate quantum spectrum.

Note that in general,  
$q_\| - {q_\|}^t(\varphi) = l_i + q_\bot\Delta\theta_i - l
= \Delta q_\|$
for the $i^{th}$ (CAP or periodic) family while 
$q_\bot - {q_\bot}^t(\varphi) = l_i \varphi = \Delta q_\bot$ 
so that 

\be
\int d\varphi~d\q~K_s(\q,\q,t)  =  \int d\varphi dq_\| dq_\bot 
{\sin(\kt \Delta q_\|) \over \pi \Delta q_\|}
{\sin(\kt \Delta q_\bot) \over \pi \Delta q_\bot}
\ee 

\noindent
A few approximations are now in order to keep the derivation
simple. First, we shall replace $\Delta q_\|$ in the denominator
by its mean value (at $q_\bot = 0$) so that $\Delta q_\| \simeq l - l_i$
Thus 

\be 
\int_{-w_i/ 2}^{w_i/2} dq_\bot {\sin(\kt\Delta q_\|)\over \pi \Delta q_\|}
\simeq 2  
{\sin\left(\kt\Delta\theta_i w_i/2\right)\over \kt\Delta\theta_i}
\delta_{\kt}(l - l_i)
\ee

\noindent
where $\delta_{\kt}(l - l_i) = \sin(\kt(l-l_i))/ (\pi(l-l_i))$.
We shall next consider $\kt$ finite but sufficiently large so that

\be
\int d\varphi {\sin(\kt \Delta q_\bot) \over \pi \Delta q_\bot} 
\simeq {1\over l_i}
\ee

\noindent
The $q_\|$ integration needs no approximation and yields
$\int dq_\| = l_i^p$ where $l_i^p$ refers to the length of
the orbit (primitive when $\Delta \theta_i = 0$). 
Finally then,

\be
\int d\varphi~{\rm Tr}~\Lopn_{qc}^t  \simeq 
\sum_i N{a_i\over l_i} \beta_i~\delta_{\kt}(l - l_i) 
e^{-in_i\pi - i\nu_i{\pi\over 2}}
\label{eq:classical_poly_modif}
\ee

\noindent
where $\beta_i  =  \sin(\kt\Delta\theta_i w_i/2)/
(\kt\Delta\theta_i w_i/2)$
and $a_i = w_i l_i^p$. For a polygon, $n_i$ is even for both
CAP and periodic families while $\nu_i = 0$. Thus 
as $\kt \rightarrow \infty$,
Eq.~(\ref{eq:classical_poly_modif}) reduces to Eq.~(\ref{eq:trace2}).
As before, we have neglected the contribution of isolated orbits
from both the classical and semiclassical trace formulae. 

The eigenvalues of $\Lopn_{qc}^t$ can be related to quantum
eigenvalues using Eqns.~(\ref{eq:classical_poly_modif})
and (\ref{eq:semi_poly_modif}) and we merely state the
final result :

\be 
\lambda_n^s(\varphi) = \imath \sqrt{E_n} \sin(\varphi)
\ee

\noindent
This is no different from our earlier result but in
practical terms, the use of the smoothened kernel 
justifies the use
of the $\epsilon$ neighbourhood since $K_s$ contributes 
substantially only in a small neighbourhood around $\q$, the size
of which is determined by $\kt$. Note that the correspondence
demands that the quantity $\kt$ in $K_s$ be identified with $\sqrt{E_{max}}$
where $E_{max}$ is the maximum energy that one is interested in.
Thus $\kt$ also determines the time increment in the evolution
of the kernel.  

Undoubtedly, several other smoothened kernels are just as
appropriate. For instance, if one evaluates the fraction of
trajectories in an $\epsilon$ neighbourhood of $\q$, the
hat function of width $\epsilon$ and height $1/\epsilon$ 
is the appropriate kernel to use. The role of $\kt$ is then
replaced by ${1\over \epsilon}$. On the other hand, if one chooses
a Gaussian centred at $\q$, the appropriate smoothened kernel is

\be
K_s^G(\q,\q',t) = 
{e^{-{(q_\| - {q'_\|}^t(\varphi))^2\over 2\sigma^2}} \over 
\sqrt{2\pi\sigma^2}}
{e^{-{(q_\bot - {q'_\bot}^t(\varphi))^2\over 2\sigma^2}}
\over \sqrt{2\pi\sigma^2}} 
\ee

\noindent
On performing the three ($\varphi,q_\|,q_\bot$) 
integrals similarly, it is possible to show that 

\be
\int d\varphi{\rm Tr}~K_s^G
= \sum_i {a_i\over l_i} \left\{ 1 - 
{1\over 3}\left ({w_i\Delta\theta_i \over 2^{3/2} \sigma}\right )^2 + \dots \right\}
{e^{-{(l-l_i)^2 \over 2\sigma^2}} \over \sqrt{2\pi\sigma^2} }
\ee

\noindent
which approximates Eq.~(\ref{eq:classical_poly_modif})
for orbits with $\Delta\theta_i$ small provided
$1/\sigma$ is identified with $\kt$. Thus any smoothened
delta function ought to work reasonably well. 

Finally, a few  remarks about the neglect of the isolated periodic 
orbits are in order. First, these are finite in number
and have a smaller contribution in both the (quasi)classical as
well as the semiclassical trace formula compared to periodic
and almost-periodic families. Then again, diffractive contributions
\cite {EB99} can be of the same order as isolated 
periodic orbit contributions in the semiclassical trace formula. 
Thus, in order to be able to compare the semiclassical eigenvalues 
and the eigenvalues of $\Lopn_{qc}$, we have consistently 
neglected isolated orbits.

\begin{figure}[tbp]
\psfig{figure=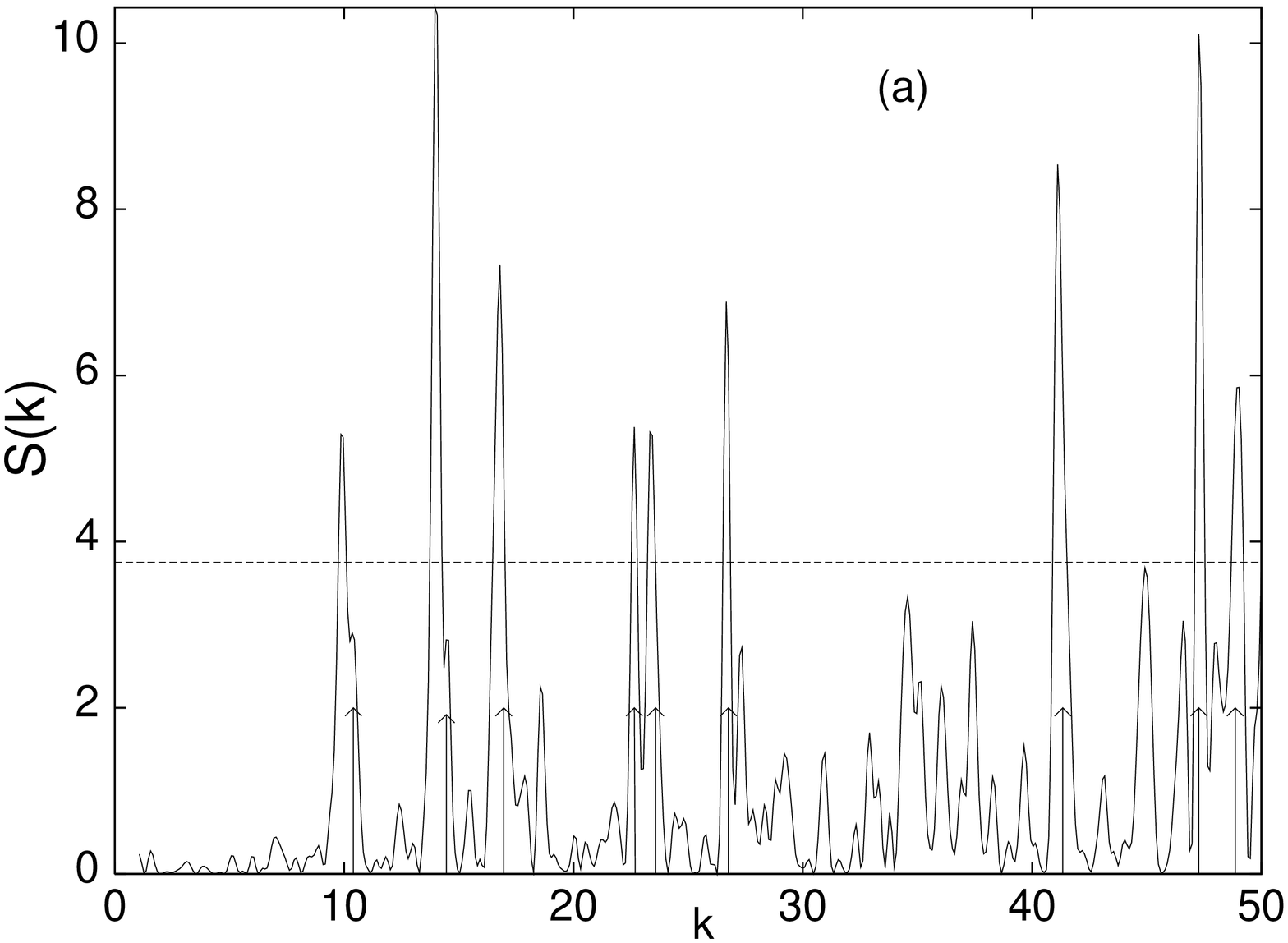,height=6cm,width=8cm,angle=270}
{\vspace*{.13in}}
\psfig{figure=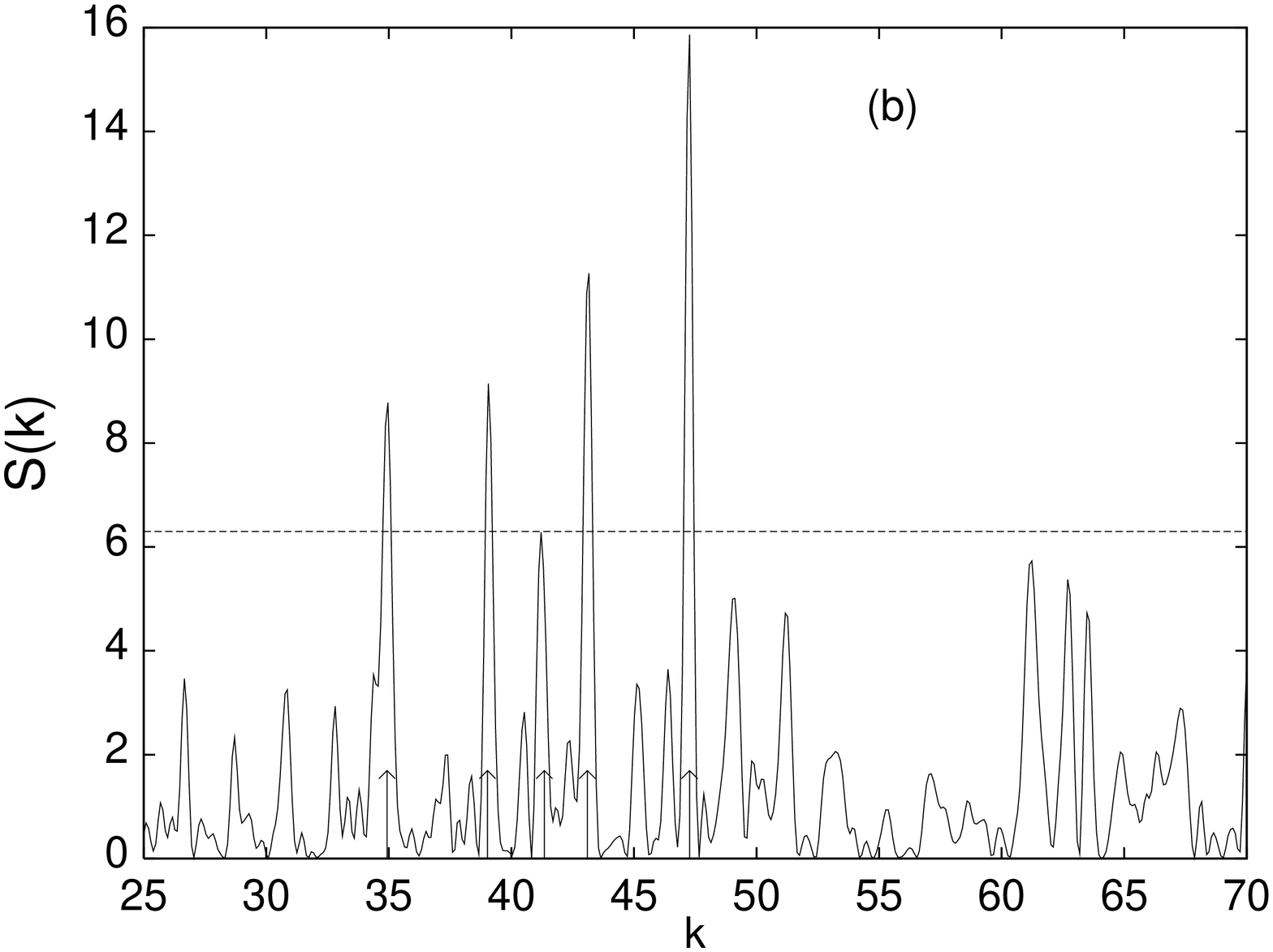,height=6cm,width=8cm,angle=270}
{\vspace*{.13in}}
\caption[ty]{The power spectrum $S(k)$ of $G(t)e^{-\beta t^2}$ as
a function of $k$ is shown for two different sets of points
$(\q,\q'=\q)$ in (a) and (b) respectively. The arrows in (a) 
mark the exact quantum eigenenergies corresponding to the
nine tallest peaks while those in (b) mark the five tallest
peaks. The dashed lines are used to mark the demarcate the 
tallest peaks in both cases. Here $\kt = 100$, $\beta = 0.01$
and the length of each trajectory is $2048\pi/\kt$.
There exist 37 eigenvalues for $k < 50$. } 
\label{fig:4}
\end{figure}

We provide some numerical results now. We first evaluate
the function $G(t) = \int d\varphi~\tilde{K}_{qc}$ for the 
($199\pi/1011,31\pi/103$) triangle of unit base length
using about 10000 trajectories for the $\varphi$ integration.
To achieve smoothening, we use the Hanning window function together
with a Gaussian damping and find the power spectrum, $S(k)$, of 
$G(t) e^{-\beta t^2}$. This is the intensity 
weighted spectrum, $\sum_n |\phi_n(\q)|^2 \delta_s(k-k_n)$ (where
$\delta_s$ is a smoothened delta function),  
for a single pair of points ($\q,\q'=\q$).
Thus, the number of distinct and unambiguous peaks in $S(k)$
depends on the value
of $\kt$ and $|\phi_n(\q)|^2$ so that by changing $\q$, 
a different set of peaks may be generated (see fig.~\ref{fig:4}). 
In fig.~\ref{fig:4}a, the nine highest peak locations (those
above the dashed line) are compared 
with the exact quantum eigenenergies while in fig.~\ref{fig:4}b, 
the highest five are compared. In all cases, the agreement 
is good and the slight difference between peak locations
and the exact quantum levels is due to the semiclassical
nature of the calculation.

The intensity weighted spectrum of fig.~\ref{fig:4} 
thus demonstrates that arbitrary
trajectories can be used to extract information about the quantum
eigenstates. Can this however be used as an effective quantization method
to determine all eigenvalues ? Fig.~\ref{fig:3} indicates that this 
may indeed be possible. A natural way of achieving this is by integrating
the intensity weighted spectrum over ${\bf q}$ \cite{average}.
In practice, only a
few ${\bf q}$ values (typically 10-20) provide information of 
about 80 \% of the eigenvalues in a given energy range.

\section{Discussion and conclusions}
\label{sec:conc}

It is worth recalling the sources of error in the
quantization recipe presented here. The theoretical
basis presented in sections \ref{sec:review} and \ref{sec:new}
clearly indicates that the eigenvalues obtained using 
this method are at best ``semiclassical'' in nature since
the modified trace formula, which connects the eigenvalues of
$\Lopn_{qc}$ with the quantum eigenvalues, is only approximate
as higher order corrections (${\cal O}(1/k)$ 
due to isolated and diffractive orbits) have been neglected. In cases 
when the corrections are zero (such as in the rectangular
or equilateral billiards), the ATQM does give exact results.
However, there are examples of other integrable billiards
(such as the circle billiard) when corrections to the 
semiclassical trace formula are non-zero and the ATQM
gives only the EBK eigenvalues \cite{keller}.  

There can however be a further source of error in the approximation
$N_p \simeq N$ (see the discussion after Eq.~\ref{eq:trace2})
especially in non-generic systems where a significant
fraction of long periodic orbits access fewer momentum directions
than the permissible number. However, we believe that in
generic situations, the approximation is fair.

Finally, since a comparison of this method with periodic
orbit theory is inevitable, it must be reiterated that the
``usual'' periodic orbit theory neglects closed almost-periodic
orbits and hence cannot give correct results in generic
situations. The modified periodic orbit theory (or trace formula)
does include CAP orbits. However, due to difficulties in
enumerating them, energy domain quantization is expected to
be quite cumbersome while the ATQM scores well due to its simplicity.
There are {\em non-generic} situations however
where closed almost periodic trajectories do not contribute
significantly and this occurs in systems where the number of
momentum directions accessible is small and $\Delta\theta_i$ is
therefore always ``large''. In such situations, diffraction
effects assume greater significance and the ATQM may not be
as effective.

In summary, we have provided a simple algorithm for 
determining the intensity weighted 
semiclassical spectrum of polygonal billiards and shown that
a smoothened delta function kernel for the quasiclassical
evolution of densities is necessary at finite energies 
to incorporate the 
effects of closed almost-periodic orbits.

\newcommand{\PR}[1]{{Phys.\ Rep.}\/ {\bf #1}}
\newcommand{\PRL}[1]{{Phys.\ Rev.\ Lett.}\/ {\bf #1}}
\newcommand{\PRA}[1]{{Phys.\ Rev.\ A}\/ {\bf #1}}
\newcommand{\PRB}[1]{{Phys.\ Rev.\ B}\/ {\bf #1}}
\newcommand{\PRD}[1]{{Phys.\ Rev.\ D}\/ {\bf #1}}
\newcommand{\PRE}[1]{{Phys.\ Rev.\ E}\/ {\bf #1}}
\newcommand{\JPA}[1]{{J.\ Phys.\ A}\/ {\bf #1}}
\newcommand{\JPB}[1]{{J.\ Phys.\ B}\/ {\bf #1}}
\newcommand{\JCP}[1]{{J.\ Chem.\ Phys.}\/ {\bf #1}}
\newcommand{\JPC}[1]{{J.\ Phys.\ Chem.}\/ {\bf #1}}
\newcommand{\JMP}[1]{{J.\ Math.\ Phys.}\/ {\bf #1}}
\newcommand{\JSP}[1]{{J.\ Stat.\ Phys.}\/ {\bf #1}}
\newcommand{\AP}[1]{{Ann.\ Phys. (N.Y.)}\/ {\bf #1}}
\newcommand{\PLB}[1]{{Phys.\ Lett.\ B}\/ {\bf #1}}
\newcommand{\PLA}[1]{{Phys.\ Lett.\ A}\/ {\bf #1}}
\newcommand{\PD}[1]{{Physica D}\/ {\bf #1}}
\newcommand{\NPB}[1]{{Nucl.\ Phys.\ B}\/ {\bf #1}}
\newcommand{\INCB}[1]{{Il Nuov.\ Cim.\ B}\/ {\bf #1}}
\newcommand{\JETP}[1]{{Sov.\ Phys.\ JETP}\/ {\bf #1}}
\newcommand{\JETPL}[1]{{JETP Lett.\ }\/ {\bf #1}}
\newcommand{\RMS}[1]{{Russ.\ Math.\ Surv.}\/ {\bf #1}}
\newcommand{\USSR}[1]{{Math.\ USSR.\ Sb.}\/ {\bf #1}}
\newcommand{\PST}[1]{{Phys.\ Scripta T}\/ {\bf #1}}
\newcommand{\CM}[1]{{Cont.\ Math.}\/ {\bf #1}}
\newcommand{\JMPA}[1]{{J.\ Math.\ Pure Appl.}\/ {\bf #1}}
\newcommand{\CMP}[1]{{Comm.\ Math.\ Phys.}\/ {\bf #1}}
\newcommand{\PRS}[1]{{Proc.\ R.\ Soc. Lond.\ A}\/ {\bf #1}}


\begin{references}
\bibitem{explain_CAP} These orbits 
close in position but the angle, $\Delta\theta$, between the 
initial and final momentum (at the point where the orbit
closes) is non-zero but small. In contrast, $\Delta\theta = 0$
for periodic families.
\bibitem{circle} For the circle billiard, $\varphi$ is related to the 
angular momentum rather than the linear momentum.
\bibitem{asymp} The function $f(x)$ is the asymptotic expansion of
$J_0(x)$.
\bibitem{epsilon_arbit} The cell size, $\epsilon$, is arbitrary
at this point and appears to be a free parameter. We shall relate 
$\epsilon$ to $E_{max}$ where $E_{max}$ is the maximum energy that one 
is interested in.
\bibitem{adhoc} The exact choice of the kernel is dictated by the 
modified semiclassical trace formula. There are however other 
options as discussed in the text.
\bibitem{average} Averaging over $q$ as in fig.~\ref{fig:3} is a crude way of
extracting this information.
\end{references}
\end{document}